\documentclass[aps,pra,twocolumn,amsmath,amssymb,letterpaper,groupaddresses,superscriptaddress,10pt]{revtex4}
\usepackage[FIGTOPCAP]{subfigure}
\usepackage[final]{graphicx}

\usepackage{times}
\usepackage{braket}
\usepackage{latexsym}
\usepackage{graphicx}
\usepackage{verbatim,times,bbm}
\usepackage{color}
\usepackage{appendix}
\usepackage{bm}
\usepackage{bbm}
\usepackage{amsmath}
\usepackage{float}

\def\uno{\mbox{1 \kern-.59em {\rm l}}}

\def\be{\begin{equation}}
\def\ee{\end{equation}}
\def\ba{\begin{eqnarray}}
\def\ea{\end{eqnarray}}

\def\la{\langle}
\def\ra{\rangle}

\begin{document}
 
\title{Critical slowing down of multi-atom entanglement by Rydberg blockade}

\author{Tahereh Abad}
\affiliation{Department of Physics, Sharif University of Technology, Tehran, Iran}
\affiliation{Department of Physics and Astronomy, Aarhus University, 8000 Aarhus C, Denmark}
\author{Klaus M\o lmer}
\affiliation{Department of Physics and Astronomy, Aarhus University, 8000 Aarhus C, Denmark}

\begin{abstract}
Laser excitation pulses that lead to perfect adiabatic state transfer in an ensemble of three-level ladder atoms lead to highly entangled states of many atoms if their highest excited state is subject to Rydberg blockade. Solution of the Schr\"odinger equation shows that it is increasingly difficult to ensure the adiabatic evolution as the number of atoms increases. A diminishing energy gap, significant variations in collective observables, and increased work fluctuations link the critical slowing down of the adiabatic evolution with a quantum phase transition-like behavior of the system.
\end{abstract}

\maketitle

\section{Introduction}\label{intro}
Adiabatic passage, where a quantum system follows the instantaneous eigenstates of a time varying Hamiltonian plays a key role in atomic and molecular spectroscopy \cite{Vitanov},  as well as in schemes to prepare entangled states \cite{Chaudhury} and in proposals for quantum computing and simulation \cite{Aharonov}. Deviations from adiabaticity occur and can be partly mitigated by optimal control or explicit shortcut-to-adiabaticity methods that involve additional driving Hamiltonians to counteract non-adiabatic couplings \cite{Demirplak, Berry, Torrontegui, Motzoi}. The ability to follow an adiabatic eigenstate of a quantum system is related to how rapidly the state changes and how large is the energy gap with the other eigenstates. For macroscopic systems we may observe a critical slowing down if the ground state of the system undergoes a quantum phase transition (QPT) \cite{Sachdev, Zurek}.

In this article, we revisit a theoretical protocol \cite{Klaus} to prepare entangled states of a number of three-level atoms subject to the stimulated Raman adiabatic passage (STIRAP) laser pulse excitation scheme \cite{Vitanov}. All atoms start in their ground state, which is a (dark) eigenstate of the system when a resonant laser is applied to couple the unoccupied middle and upper states, see Fig.\ref{pulse}. The gradual application of a resonant laser coupling of the ground and middle states transforms the dark state into a superposition of the ground and upper excited states. The dark state eventually becomes the excited states, when the laser coupling the upper states is finally switched off. If the upper state is a Rydberg state, and if the atoms are close enough to experience a significant interaction between Rydberg excited states, however, the laser fields are not resonant with the ultimate excitation of all atoms into the Rydberg product state, and an ensemble of several atoms instead explores a family of entangled states. Instead of being all excited at the end of the STIRAP pulse sequence, the atoms end up in a state with exactly the same number of atoms in the two lower states and a single atom (no atom) occupying  the Rydberg state if the number of atoms is odd (even) \cite{Klaus}. By inclusion of an ancillary ground state, it is possible to exploit the dependence on the parity of the atom number to generate GHZ superposition states with all atoms occupying one and the other ground states \cite{Klaus}.

For applications, it is crucial to perform the adiabatic excitation process fast, but we find that the scheme slows down as the number of atoms increases. While the system retains a collective (dark) atomic eigenstate through the entire evolution, for many atoms its energy gap with other states becomes vanishingly small and the collective properties of the state changes character during the application of the STIRAP pulses.  We thus have a situation very similar to the one encountered in quantum phase transitions in many-body systems. While our system does not have a thermodynamics limit in the usual sense, the purpose of the present article is to show that a number of criteria for quantum phase transitions and critical phenomena match our system and explain aspects of the observed dynamics.

The article is organized as follows. In Sec. \ref{Sec2}, we present the physical model, and we discuss the structure of the adiabatic eigenstates and the energy spectrum. In Sec. \ref{Sec3}, we present numerical solutions of the time dependent Schr\"odinger equation, and we introduce different quantities that illuminate the non-adiabaticity from the perspective of a quantum phase transition. In Sec. \ref{Sec4}, we show that a number of quantities, introduced in the literature to distinguish quantum phase transitions are equivalent, while, indeed, they illustrate different properties of the dynamics. Sec. \ref{Sec5} concludes the article.

\section{STIRAP process towards a quantum correlated state}\label{Sec2}

\subsection{Effective Hamiltonian and adiabatic eigenstates}\label{Sub1Sec1}
Following \cite{Klaus}, we consider an ensemble of $N$ three-level atoms, with ground state $\ket{g}$, excited state $\ket{e}$ and Rydberg excited state $\ket{r}$ subject to STIRAP excitation pulses. The atoms are all initially prepared in the ground state, and due to identical laser-atom interaction, their collective state is symmetric under permutations and we can employ the basis of collective states $\ket{n_g,n_e,n_r}$  characterized by the collective occupation of the three internal states in the description of the system.

Using bosonic creation and annihilation operators $a_{i}^{(\dagger)}$, $i=g,e$, the laser coupling of the lower levels can be rewritten in a collective spin description ($\hbar=1)$,
\be
H_{J_{x}}(t)= -\Omega_{1}(t) J_{x}=-\frac{1}{2}\Omega_{1}(t)(a_{g}^{\dagger}a_{e}+a_{g}a_{e}^{\dagger}).
\ee
Atoms within $\sim 10\ \mu$m separation experience strong dipolar interactions among Rydberg excited states that shift the energy levels enough to prevent resonant excitation of more than a single atom. This restricts the values of the occupation number $n_r$ to $0$ and $1$, and introducing collective Pauli operators, $\sigma^+$ and $\sigma^-$, raising and lowering the value of $n_r$ by unity, we can write the laser coupling of the upper atomic levels as an effective Jaynes-Cummings (JC) Hamiltonian,
\be
H_{JC}(t)=-\frac{1}{2}\Omega_{r}(t)(a_{e}\sigma^{+}+a_{e}^{\dagger}\sigma^{-}).
\ee

\begin{figure}[htbp]
\centering
\includegraphics[width=0.45\textwidth]{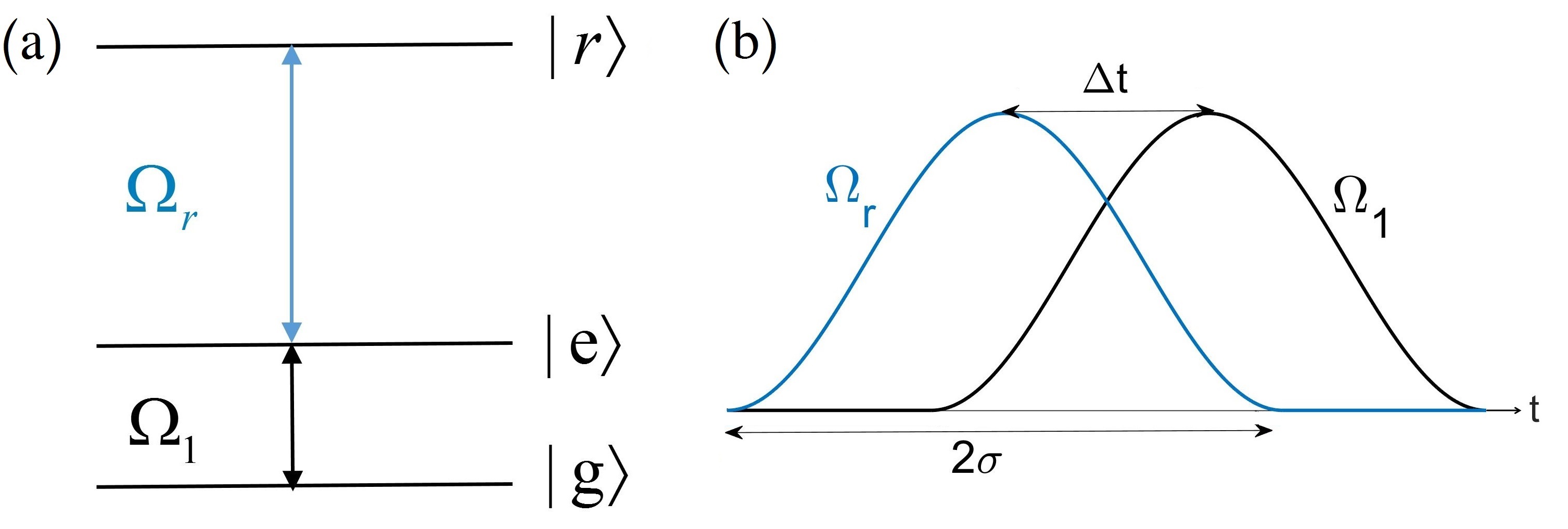}
\caption
{(a) The atomic three-level ladder system is driven by two resonant laser fields with Rabi frequencies $\Omega_{1}$ and $\Omega_{r}$ (b). The STIRAP laser pulse sequence assuming pulses, $\Omega_j(t)=\Omega_{\text{max},j}\sin^2(\frac{\pi(t-t_{sj})}{2\sigma})$ for $t_{sj} <t< t_{sj}+2\sigma$, with $2\sigma$ duration and $\Delta t = t_{s1}-t_{sr}$ peak separation.}
\label{pulse}
\end{figure}

The system is subject to the time dependent total Hamiltonian
\be \label{H}
H(t)=H_{JC}(t)+H_{J_{x}}(t),
\ee
which for all times has a spectrum of eigenvalues that is symmetric around the zero energy dark state.

\begin{figure}[h]
\centering
\includegraphics[width=0.47\textwidth, height=4.2 cm]{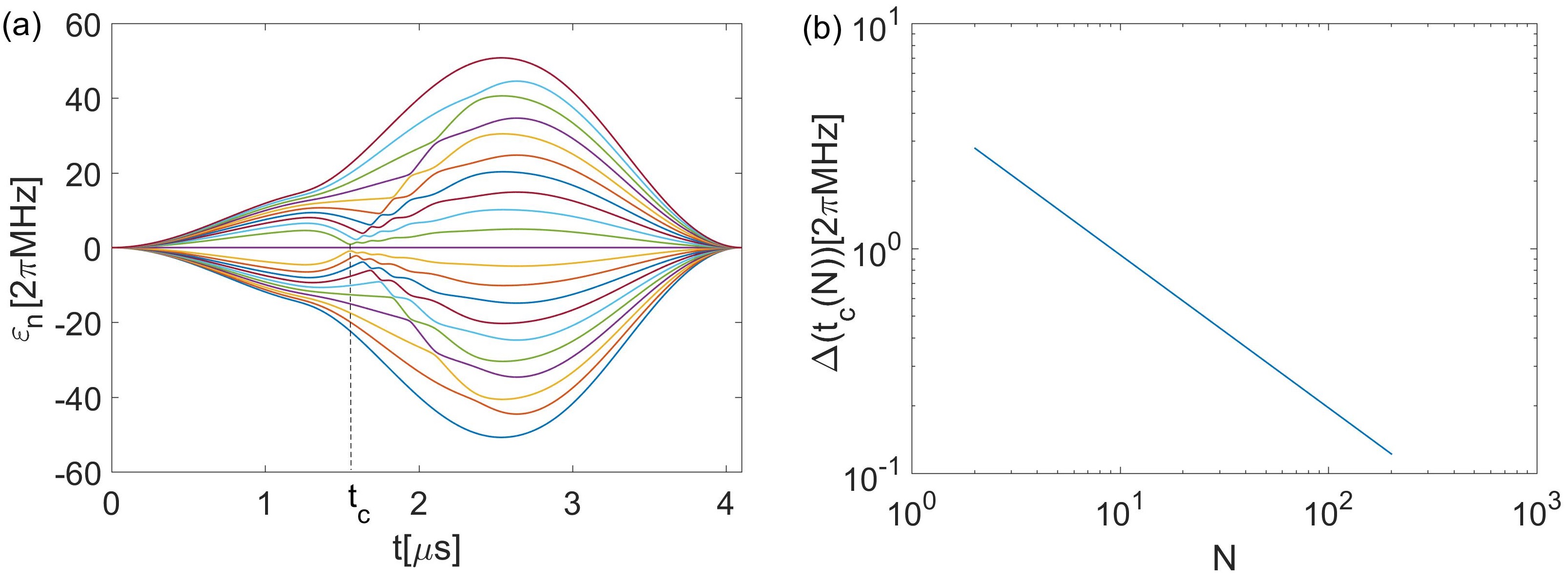}
\caption{(a) The energy spectrum of the time dependent Hamiltonian (\ref{H}) is shown for $N=10$ atoms assuming the parameters $\Omega_{\text{max},1}=\Omega_{\text{max},r}=2\pi\times10\text{MHz}$. $\bigtriangleup t=1.1 \mu s$ and $\sigma=1.5 \mu s$. The minimum energy gap is obtained at $t_c$ is $1.53 \mu s$.  (b) The local minimum energy gap between the zero energy and the nearest non zero energy eigenvalues scales as $\frac{1}{\sqrt{N}}$.}
\label{eig}
\end{figure}

Fig.\ref{eig} shows the energy spectrum for $N=10$ atoms, subject to the time dependent pulses shown in Fig.\ref{pulse}(b). At early times, only the upper atomic levels are coupled according to the effective JC Hamiltonian, with energy eigenvalues $\{0,\pm \frac{1}{2}\Omega_r(t) \sqrt{n}\}$ where $n$ denotes the total number of atoms populating the two upper atomic levels, and the zero energy (dark) state is the ground product state of the atoms. At the latest times, only the lower levels are coupled according to the collective spin Hamiltonian, yielding an equidistant energy spectrum. Here, the dark state has evolved into the highly entangled collective $J_x = 0$ eigenstate. This state has useful properties for spectroscopy and its entanglement can be assessed by measurement of only collective properties \cite{Klempt}.

When both laser fields are on, the eigenstates are not analytically known, but we see that they are continuously connected, and in particular, the adiabatic passage permits preparation of the highly entangled $J_x=0$ state from a simple product state. We also observe, however, that despite both separate eigenspectra showing increased level spacing with the Rabi frequencies $\Omega_{1,r}$, the joint action of both Hamiltonians conspires to form a contraction of the energy eigenvalues towards the zero energy dark state at a critical time $t_c$ during the time evolution. The JC and the collective spin Hamiltonians both have energy levels around zero that are independent on the atom number $N$, but our numerical diagonalization yields an energy gap between the zero energy and the next non zero energy, $\Delta=\Delta(t,N)$ which decreases at the critical time $t_c$ as
\be
\Delta(t_c(N))\sim N^{-\frac{1}{2}},
\ee
as shown in Fig.\ref{eig}.

In many-body physics, one conventionally defines the thermodynamic limit as the limit where the number of particles goes to infinity, assuming a constant density and unchanged local interactions between particles. In this paper we will study similarities between the large $N$ behavior of our system and quantum phase transitions, but we note that we need an increasing density for all our atoms to be within reach of the Rydberg interaction distance and interact equally strongly with each other. This poses a fundamental difference, e.g., in the distinction of intrinsic and extrinsic properties between our systems and many-body systems with local interactions. The purpose of the article is to investigate which properties of our system are similar to the ones characterizing true phase transition dynamics.

\section{Solution of the time dependent Schr\"{o}dinger equation}\label{Sec3}
\begin{figure}[htbp]
\centering
\includegraphics[width=0.30\textwidth, height=3.5cm]{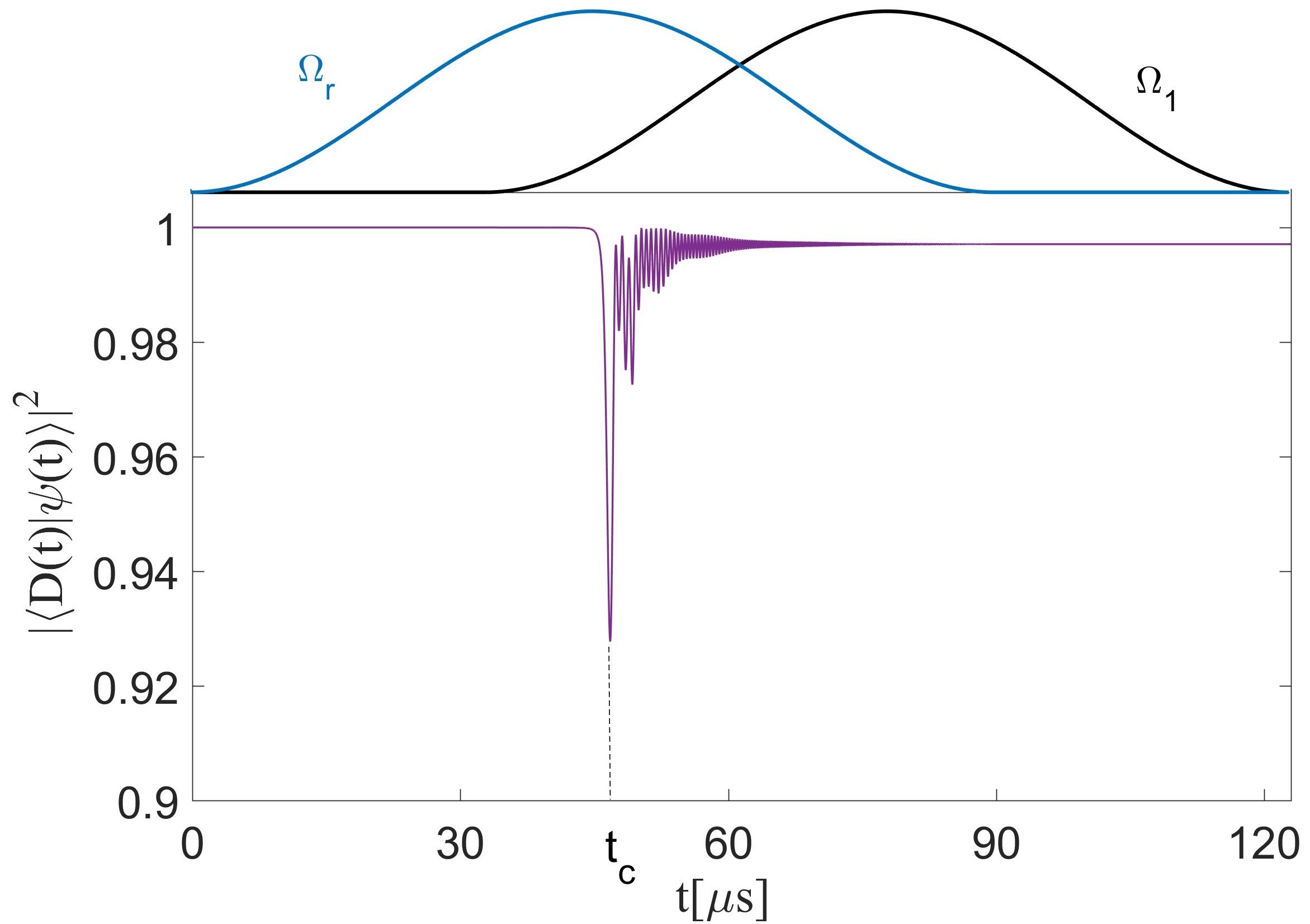}
\caption{The overlap between the solution of the time-dependent Schr\"odinger equation, $\ket{\psi(t)}$, and the zero energy adiabatic state, $\ket{D(t)}$, for $N=10$. We assume the pulse shapes specified in Fig.2, but the longer duration $\sigma=45 \mu$s and $\Delta t = 33\mu$s.}
\label{CheckAd}
\end{figure}

If the system is initially prepared in the (dark) ground product state $\ket{D(t_0)} = \ket{ggg...}$, it will adiabatically follow the zero energy eigenstate $\ket{D(t)}$ of $H(t)$ if the evolution is sufficiently slow. However, changing the Hamiltonian too rapidly will cause the solution of the Schr\"{o}dinger equation  $\psi(t)$ to undergo non-adiabatic transitions to other eigenstates. In Fig.\ref{CheckAd} we show the time dependent overlap fidelity, $F=|\la \psi(t)|D(t)\ra|^2$ for $N=10$, obtained for a total duration of the STIRAP pulses of $120 \mu s$. 

\begin{figure}[htbp]
\centering
\includegraphics[width=9.50cm, height=4.0cm]{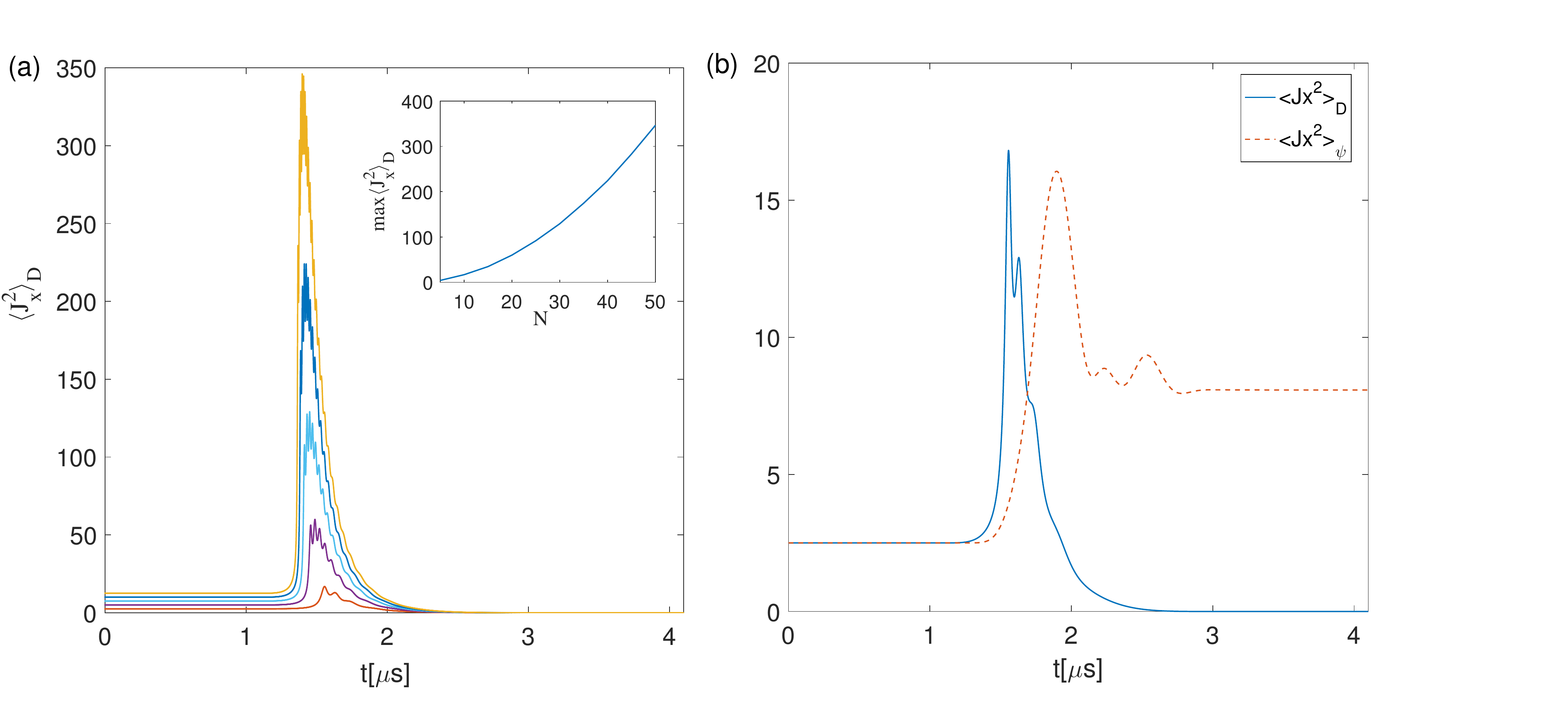}
\caption{(a) The mean value of $J_x^2$ in the zero energy adiabatic eigenstate is shown as a function of the time argument $t$ in the Hamiltonian (3). The curves from bottom to top, correspond to $N=10, 20, 30, 40$ and $50$. The inset shows the quadratic behavior of the maximum values as a function of $N$. (b) The dashed red curve and the solid blue curve show a comparison of the mean value of $J_x^2$ for a solution of the time dependent Schr\"odinger equation and the adiabatic eigenstate for $N=10$.
}
\label{Jx2D}
\end{figure}

We will focus our attention on the dynamics around $t_c$ where the difficulty to follow the adiabatic eigenstates has two causes: the reduced energy gap makes the other low lying eigenstates energetically accessible to the system in the presence of even weak coupling, and the eigenstates rapidly change their physical character (causing, indeed, the non-adiabatic coupling). We illustrate the rapid variation in the eigenstates by the variance of the collective $J_x$ observable in the dark adiabatic eigenstate in Fig.\ref{Jx2D}(a). Curves are shown for different values of $N$, all starting at values proportional with $N$ in the product ground state, reaching maximal values that scale as $N^2$, and returning to a vanishing value in the final $J_x=0$ eigenstate. In addition to showing the dramatic change as the systems is evolved, it emphasizes another defining property of quantum phase transitions, namely the change in macroscopic behaviour, as has been recently quantified by the fluctuations of collective observables \cite{Tahereh, Mazieroa, Hauke, Jeong}. For a general definition of macroscopicity one considers the maximally fluctuating quantity over all sums of single particle operators $A=\sum^N_{i=1} A_i$, where $A_i$ has eigenvalues $\pm 1$, and one introduces the concept of an effective size
 \begin{equation}\label{effSize}
N_{\text{eff}}(\psi)=\max_{A}\text{Var}(A)/N.
 \end{equation}
where $\text{Var}(A)=\langle \psi| A^2 |\psi \rangle-\langle \psi| A |\psi \rangle^2$. If $N_{\text{eff}}={\cal O}(N)$ , we have a macroscopically correlated state while, if $N_{\text{eff}}={\cal O}(1)$, correlations only manifest themselves at the microscopic level of few particles.

While the variance of $J_x$ goes to zero, the two orthogonal spin components acquire macroscopic fluctuations in the $J_x=0$ eigenstate where
\be
\langle J_y^2 \rangle+\langle J_z^2 \rangle= J(J+1).
\ee
where $J_{y}=\frac{i}{2}(a_{g}a_{e}^{\dagger}-a_{g}^{\dagger}a_{e})$, $J_{z}=\frac{1}{2}(a_{e}^{\dagger}a_{e}-a_{g}^{\dagger}a_{g})$ and $J=\frac{N}{2}$.
Our system does not offer the usual distinction between intrinsic and extrinsic many-body properties, and unlike many-body systems that display macroscopicity only around singular phase transition points, our system has no length scale and permits correlations among all particles that lead to macroscopic fluctuations of the total spin components for a range of Hamiltonian parameters.

The inability to follow the adiabatic eigenstate around $t_c$ shown in Fig.(\ref{CheckAd}) can be understood as a consequence of not being able to establish the macroscopic fluctuations fast enough. This is illustrated in Fig.\ref{Jx2D}(b), which shows the time dependent variance of $J_x$ (the dashed red curve) during evolution of the quantum system under STIRAP pulses with finite duration. For $N=10$ atoms one observes how the increase in the variance lags behind the one of the adiabatic eigenstate, and also how the variance does not reach the vanishing final value after the STIRAP pulse sequence.

\section{Different characteristics of a quantum phase transition}\label{Sec4}
In this section we discuss how the deviation from adiabatic evolution of the quantum system is connected with the critical slowing down of quantum systems near quantum phase transitions. A number of different measures to characterize quantum phase transitions can be connected to the observed properties of our system.

\subsection{Rapid variation of the adiabatic eigenstates}\label{Sec4a}
Consider the instantaneous eigenvalues $\varepsilon_n(t)$ and eigenstates $\ket{n(t)}$ of the time-dependent Hamiltonian (\ref{H}). We are particularly interested in the dark state $\ket{D(t)}:=\ket{{\bf 0}(t)}$ with zero energy, $\varepsilon_{\bf 0}(t)=0$, and which is readily prepared as the ground product state at the beginning of the STIRAP pulse sequence. If the Hamiltonian is changed in a time dependent manner, also the instantaneous eigenstate changes, and to quantify the magnitude of non-adiabatic coupling out of the dark state, we introduce the rate of change,
\be \label{R}
R(t):=\Vert |\partial{\bf 0}(t) \rangle \Vert,
\ee
where $|\partial{\bf 0}(t) \rangle \equiv \frac{\partial}{\partial t}\ket{{\bf 0}(t)}$ and $\Vert \Psi \Vert^2=\langle \Psi|\Psi\rangle$.
Like the rapid change of properties of the ground state of many-body systems near a quantum phase transitions, we show in Fig.\ref {WorkF}(a) that the rate of change of our dark state diverges in the large $N$ (thermodynamic) limit at a critical moment during the evolution.

\subsection{Increased work fluctuation under counterdiabatic driving}\label{Sec4b}
Following \cite{Demirplak, Berry}, a technique called  counterdiabatic driving (CD) has been shown to drive the quantum evolution of a system exactly along the instantaneous eigenbasis $\ket{n(t)}$ of $H_{0}(t)$, by applying an additional Hamiltonian
\be \label{H1}
H_{1}=i\hbar\sum_{n}(\ket{\partial_{t}n}\bra{n}-\bra{n}\ket{\partial_{t}n}\ket{n}\bra{n})
\ee
to the system.
Restricting the system to populate a single eigenstate $\ket{n(t)}$, the presence of $H_{0}$ in the total Hamiltonian  $H_{CD}=H_{0}+H_{1}$ then only results in a global phase and can in principle be dropped. The shortcut to adiabaticity (STA) method suppresses transitions of the quantum system without the requirement of slow driving \cite{Torrontegui}. Clearly the magnitude of the Hamiltonian $H_{CD}$ depends on the desired evolution speed, and  in \cite{Campo} it was proposed that a phase transition would be characterized by increased fluctuations in the work performed by $H_{CD}$ to maintain the instantaneous eigenstate of a many-body system,
 \be \label{Work}
 \delta(\Delta W(t))^2:=\text{Var}[W(t)]-\text{Var}[W(t)]_{\text{ad}},
 \ee
where $\text{Var}[W(t)]=\text{Var}(H_{CD})$ and $\text{Var}[W(t)]_{\text{ad}}=\text{Var}(H_{0})$.

Suppose that we evolve the system along the zero energy eigenstate $\ket{{\bf 0}(t)}$ with $\text{Var}_{\ket{{\bf 0}}}(H_{0})=0$. To calcuate $\text{Var}_{\ket{{\bf 0}}}(H_{CD})$, we apply Eq.(\ref{H1}) which leads to $\langle{\bf 0}|H_{CD}|{\bf 0}\rangle=0$ and
 \ba
 \langle {\bf 0}|H_{CD}^2|{\bf 0} \rangle&=&-\langle {\bf 0}|(\sum_{n}\ket{\partial_{t}n}\bra{n})^2|{\bf 0}\rangle+\langle {\bf 0}|\partial_{t}{\bf 0}\rangle^2 \nonumber \\
 &=&\sum_{n\neq {\bf 0}}\langle\partial_{t}{\bf 0}|n\rangle\langle n|\partial_{t}{\bf 0}\rangle,
 \ea
where, in the last equality, we use $\langle {\bf 0}|\partial_{t}n\rangle=-\langle {\partial_{t}\bf 0}|n\rangle$. Due to reality of $H_0$, (\ref{H}), we have $\langle\partial_{t}{\bf 0}|{\bf 0}\rangle=\langle{\bf 0}|\partial_{t}{\bf 0}\rangle=0$, and the work fluctuation is given by
 \be \label{WorkRate}
 \delta(\Delta W(t))^2=\text{Var}(H_{CD})=\Vert |\partial_{t}{\bf 0}(t) \rangle \Vert^2,
 \ee
We see that the fluctuations of the work done on or by the system during the CD protocol is given by $R(t)^2$ defined in Eq.(\ref{R}).
\begin{figure}[htbp]
\centering
\includegraphics[width=9.50cm, height=4.0cm]{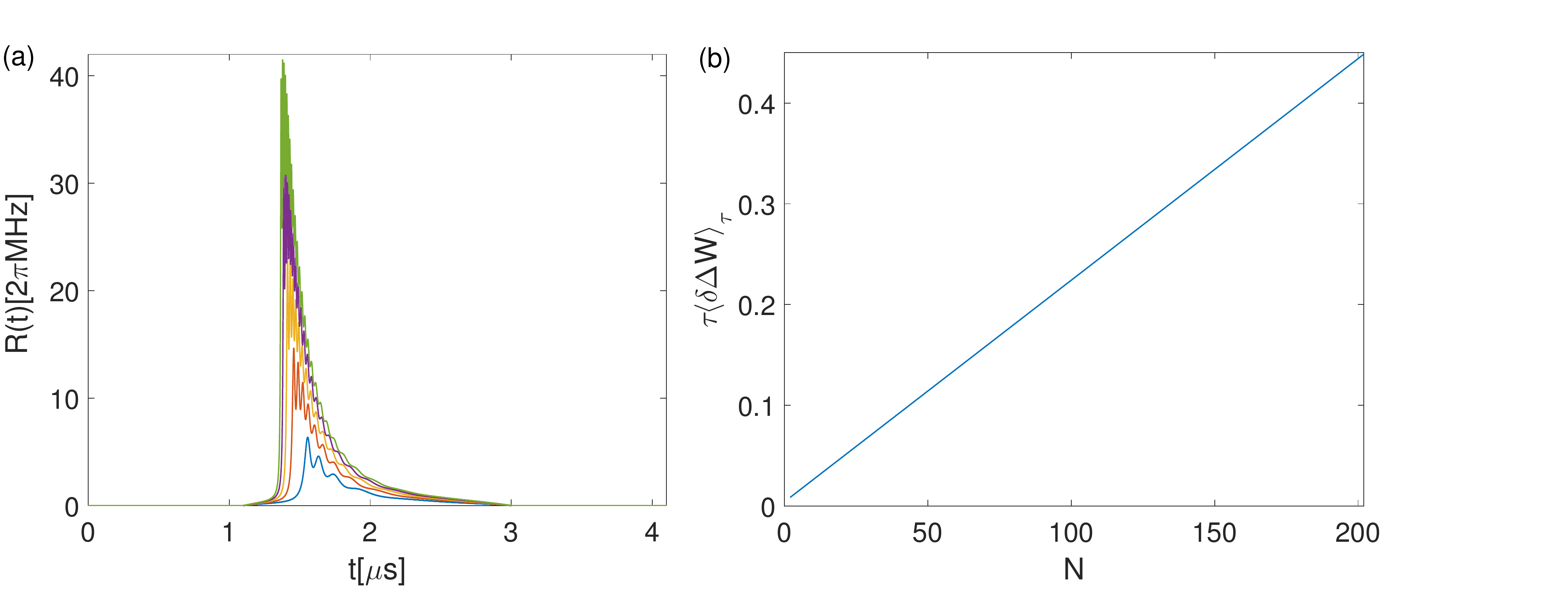}
\caption{(a) During the evolution, the rate of change of the adiabatic eigenstate (\ref{R}), equivalent to the square root of instantaneous work fluctuations (11) of the counterdiabatic driving Hamiltonian, exhibits a pronounced peak in the neighborhood of a critical point. The curves from bottom to top, correspond to $N=10, 20, 30, 40$ and $N =50$. (b) For a fixed duration,$\tau=4.1 \mu s$, of the STIRAP scheme,  the time-averaged work fluctuations (\ref{alpha}) scale linearly with the system size $N$.}
\label{WorkF}
\end{figure}

In Ref.\cite{Campo} it was found that during the STA dynamics induced by CD, the time averaged work fluctuations $\langle \delta \Delta W\rangle_{\tau}$ exhibit a universal scaling
\be \label{alpha}
\tau \langle \delta \Delta W \rangle_\tau = \int_0^{\tau} \langle \delta \Delta W(t) \rangle dt \sim N^\alpha,
\ee
where $\delta\Delta W(t) =\sqrt{\delta(\Delta W(t))^2}$, and the exponent $\alpha$ depends on the dimensionality, the size and the correlation length of the system.

Such a scaling is shown in Fig. \ref{WorkF}(b), where a fit to the numerical data leads to the power-law exponent $\alpha=1$. As we have no effects of spatial dimensionality or finite correlation lengths, we shall not insist on the precise interpretation of $\alpha$, but merely note that our system obeys a scaling with system size of the work fluctuations. This is consistent with the need for the counterdiabatic driving Hamiltonian to establish the macroscopic spin fluctuations, shown in Fig.4.

\subsection{Neighboring state fidelity}\label{Sec4c}
Another way to analyze QPT behavior \cite{Zanardi1,Zanardi2} compares ground states corresponding to slightly different values of the Hamiltonian through the state fidelity
\be \label{fid}
{\cal F}(t,\epsilon):=|\langle\psi(t-\epsilon)|\psi(t+\epsilon)\rangle|.
\ee
Indeed, Eq. (\ref{fid}) is expected to decrease abruptly at quantum phase boundaries where a small change of the Hamiltonian yields a dramatic ground state variation.

\begin{figure}[htbp]
\centering
\includegraphics[width=0.77\textwidth, height=4cm]{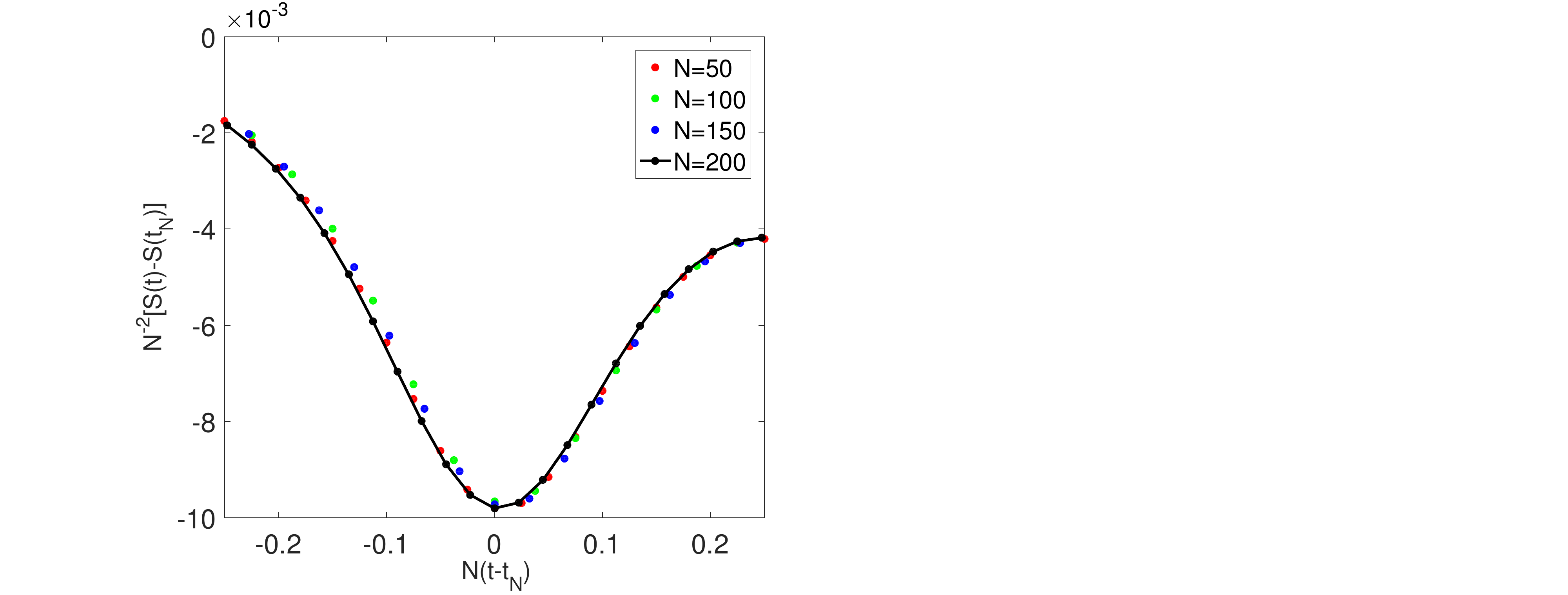}
\caption{Finite size scaling of the state fidelity near the energy level avoided crossing. The figure shows that the function ${\cal Q}(N(t-t_N))=N^{-2}(S(t,N)-S(t_N,N))$ has almost the same value for $N=50, 100$ and $N =150$, as function of the shifted and rescaled time $N(t-t_N)$.}
\label{Fidelity}
\end{figure}

The first non-zero order of the Taylor expansion of the overlap function
\be
S(t):=\partial^2_\epsilon {\cal F}(t,\epsilon)|_{\epsilon=0}
\ee
is clearly closely connected to the rate of change of the adiabatic eigenstate $R(t)$,and it is interesting to study its behavior around the avoided crossing in the eigenvalue spectrum in Fig.2. Fig.\ref{Fidelity} shows that a single function captures the variation of $S(t)$ for all values of $N$
\be
S(t,N)-S(t_N,N)=N^{2}{\cal Q}(N(t-t_N)),
\ee
where $t_N$ yields the minimum value of $S(t)$ for each value of $N$ and ${\cal Q}$ is a universal function of $N(t-t_N)$, derived numerically and shown in Fig.\ref{Fidelity}. This is reminiscent of the $N^{1/\nu}(t-t_m)$  scaling behavior \cite{Barber} of diverging observables in second order quantum phase transitions, where we shall again not assign too much interpretation to the exponent $\nu$ being unity, as we do not have an extended system with clearly defined intrinsic and extrinsic variables.

\section{Conclusion}\label{Sec5}

In this article we have studied a scheme for preparation of multi-atom entangled states by adiabatic passage. Despite the initial and final Hamiltonians having both well separated energy eigenstates, their weighted sum at a crucial moment during the time evolution features a diminishing energy gap and system eigenstates that change rapidly so that it becomes difficult to maintain adiabaticity and reach the desired final state. A similar collapse of the energy spectrum occurs in several physical models, for example the Jaynes-Cummings model subject to a classical drive field  \cite{alsing}, and while it is attractive to use adiabatic passage, its feasibility as one explores larger system sizes and state spaces is an important concern.

We have explored different means to characterize the lack of adiabaticity in our system and we have found multiple similarities with the physics of phase transitions. Unlike, e.g., the Ising Hamiltonian, the interactions in our system do not depend on distance or dimensionality (we assume perfect blokcade interaction between any pair of atoms in our system). While the arguments for universal behavior and critical exponents in condensed matter systems cannot hence be directly applied to our system, the framework of phase transitions and critical dynamics provide useful insight in its dynamics.

\subsection*{Acknowledgements}
The authors acknowledge support from the Villum Foundation, and T.A. acknowledges support from the Ministry of Science Research and Technology of Iran and Sharif University of Technology under grant no. G951418. The authors thank Vahid Karimipour and Albert Benseny Cases for constructive comments.

{}

\end{document}